# DBMSs Should Talk Back Too

Yannis Ioannidis
University of Athens
Athens, Hellas (Greece)
yannis@di.uoa.gr

Alkis Simitsis
HP Labs
Palo Alto, California, USA
alkis@hp.com

## ABSTRACT

Natural language user interfaces to database systems have been studied for several decades now. They have mainly focused on parsing and interpreting natural language queries to generate them in a formal database language. We envision the reverse functionality, where the system would be able to take the internal result of that translation, say in SQL form, translate it back into natural language, and show it to the initiator of the query for verification. Likewise, information extraction has received considerable attention in the past ten years or so, identifying structured information in free text so that it may then be stored appropriately and queried. Validation of the records stored with a backward translation into text would again be very powerful. Verification and validation of query and data input of a database system correspond to just one example of the many important applications that would benefit greatly from having mature techniques for translating such database constructs into free-flowing text.

The problem appears to be deceivingly simple, as there are no ambiguities or other complications in interpreting internal database elements, so initially a straightforward translation appears adequate. Reality teaches us quite the opposite, however, as the resulting text should be *expressive*, i.e., accurate in capturing the underlying queries or data, and *effective*, i.e., allowing fast and unique interpretation of them. Achieving both of these qualities is very difficult and raises several technical challenges that need to be addressed.

In this paper, we first expose the reader to several situations and applications that need translation into natural language, thereby, motivating the problem. We then outline, by example, the research problems that need to be solved, separately for data translations and query translations.

## 1. INTRODUCTION

The role of a database management system (DBMS) installed in a particular environment is to connect its users with its underlying content. Natural language (NL) is often an important tool on both ends of this connection: on one hand, users may interact with the system in natural language, while on the other, the original database content may be in natural language. Nevertheless, for high performance and effectiveness, DBMSs prefer to deal with structured elements internally. For this reason, much research has been conducted on translating natural language queries and other user interaction expressions to (mostly) SQL, as well as on translating natural language documents to (mostly relational) database records, by extracting information from them. The above implies that the external components of a DBMS environment often talk to the DBMS in natural language. Shouldn't the DBMS also be able to talk back in the same language?

In this paper, we answer this question affirmatively and describe several applications where such translation of DBMS internal elements to natural language would be very beneficial. Obtaining such DBMS-generated natural language constructs, however, is far from trivial, making this largely-ignored problem a rich field for exploration. Recent studies advocate for –and motivate to some extent– the need for such functionality. Automating computer-to-human speech translation is recognized as one of the seven most important IT challenge for the next 25 years by Gartner analysts who examine technologies that will have a broad impact on all aspects of people's lives [3]. In this direction, we identify what we believe are the key research issues that arise and need to be addressed and some ideas for possible solutions.

Although related to some extent, translating structured database content into natural-language free text is quite distinct a problem from translating queries expressed in a formal language into equivalent commands in natural language. The former is difficult because of the size of the underlying data, the need to make choices on what to capture in a short narrative and what to ignore, and the repetitive nature of some data in the database that need to be factored out and expressed once in the text generated for effective reading by the user. The latter is difficult because the size and complexity of a query are essentially arbitrary and have no upper bounds (whereas the contents of a database necessarily follow the schema structure, which is bounded), thus allowing generation of rather convoluted queries, whose translation is a great challenge.

Any attempt towards a solution of the content and/or the query translation problems must balance out the desired concision and naturality of the generated text with the complexity of the translation process itself. Our preliminary efforts have been leading to graph representations of database contents and queries (different from one to the other), template phrases associated with parts of the graph, and then graph traversal in particular directions to compose the templates found on the way into the final text formation.

In the next section, we explore translations of database contents, for which our work has progressed further, while in the section after that, we explore translations of database queries. We conclude with some additional thoughts on the whole problem.

## 2. DATABASE CONTENTS

### 2.1 Motivation

Consider that one wants to have a textual description of the contents of a database. If the database is large, it would make sense to create a textual summary of it, otherwise, the result can be a description of the full contents. There are several situations where such translation into natural language may be useful and desirable. Creating a short company description for a business plan or a bank-loan application or collateral material for marketing are some instances. Given other appropriate schemas, one can imagine textual descriptions in several other practical cases: a short description of a museum's exhibits, possibly customized to a visitor's particular interests; a brief history of a patient's medical conditions; the highlights of a collection in a digital library, with a few sentences on the main authors in the collection; a summary of a theater play in an information portal; and others.

Whatever holds for whole databases, of course, holds for query answers as well, especially those with some nontrivial structure, i.e., entire relational databases, complex objects, and so on. Textual answers are often preferred by users, whether experienced or not, as they convey the essence of the entire query answer in an immediately understandable way. Moreover, the formation of textual answers becomes critical in all situations for people with visual impairments or reading disabilities. Using a speech recognizer [2, 7] to convert a speech signal to a query and a text-to-speech system (TTS) [7] to convert the textual form of the query answer into speech, these people would be given the chance to interact with information systems, orally pose queries, and listen to their answers.

Clearly, the idea of translating data into natural language can be extended to all other forms of primary or derived data that a database may contain. Database samples, histograms, data distribution approximations are all, in some sense, small databases and can be summarized textually as above. Describing the schema itself, its basic entities, relationships, and other conceptual primitives offered by the model it is based on, is just a special case of a database description. User profiles maintained by the system for offering personalized answers, browsing indexes, and other forms of metadata are amenable to and may benefit from natural-language translation as well.

### 2.2 Translating Database Content

Databases can be modeled conveniently as graphs, i.e., database schema graphs. The main entities, i.e., relations and attributes, constitute the nodes of the graph, whereas the relationships among them, i.e., join and projection edges, represent the edges of the graph. A projection edge, one for each attribute node, emanates from its container relation node and ends at the attribute node, representing the possible projection of the attribute in a query answer. A join edge emanates from a relation node and ends at another relation node, representing a potential join through a primary key - foreign key relationship between these relations.

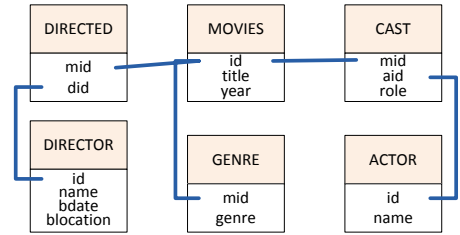

**Figure 1: Example database schema**

The graphical representation of the database schema facilitates its translation. In a previous work, we proposed a template-based method for appropriately annotating the database schema graph, and then showed how a graph traversal can produce meaningful narratives [9]. The solution suggests that both nodes and edges are annotated by appropriate template labels. These labels are assigned once, e.g., by the designer, at an initial design phase, and are instantiated at query time, in order to produce textual descriptions. Some indicating examples will be shown in the subsequent paragraphs, as the overall approach is outlined. For the full-fledged approach, we refer the interested reader to our previous work [9].

Consider for example a simple schema describing a movie database (see Fig. 1). For clarity of presentation, only relation nodes and join edges are depicted. (Without loss of generality we may assume that the names of relations and attributes are meaningful; otherwise, appropriate aliases can be used.) For translation purposes, we consider that each relation has a *conceptual* and a *physical* meaning. For example, the relation $MOVIE$ conceptually represents "movies" in real world. The physical meaning of a relation is represented by the name of one of its attributes, the one that is most characteristic of the relation tuples; this attribute is termed the *heading* attribute. In our example, $TITLE$ is the heading attribute for the relation $MOVIE$. Textual sentences about the relation contents are composed using templates based on the heading attribute, which is usually used as the subject of these sentences. For example, when we want to refer to a movie, practically, we refer to its heading attribute, the title. A template attached as a label to the projection edge connecting the relation $MOVIE$ to its attribute $YEAR$ may be "the YEAR of a MOVIE(.TITLE)". Similarly, a template attached to a join edge signifies the relationship between the heading attributes of the relations involved; e.g., "the GENRE(.GENRE) of a MOVIE(.TITLE)".

The translation of a simple database schema graph containing a single relation is performed by composing phrases based on the templates of the respective projection edges connecting the relation with its attributes. There are two alternatives for the translation of a relation's contents: (a) create a sentence using only a template based on the heading attribute of the relation or (b) using one or more templates to construct a sentence that combines the information stored in the heading attribute and the other attributes of the relation under consideration. The former results into the creation of simple phrases, such as "The director's name is Woody Allen". When we wish to combine information coming from a single relation with information coming from other relations, then, such a short version frequently seems to be adequate (more details are given below). The second alternative for translating in more detail the content of a single relation results into the creation of different clauses, one per attribute, where inevitably the same subject has to be repeated many times. To avoid this, we need to create these clauses using the appropriate templates, and then find common expressions in the clauses and replace them accordingly. For example, for a translation of relation $DIRECTOR$ involving the attributes $BDATE$ and $BLOCATION$, which store information about the birth date and birth location of a director, we may use the following templates that connect these attributes with $DNAME$, the heading attribute of the relation ('+' indicates concatenation):

$DNAME +$ " was born" $+$ " in " $+ BLOCATION$

$DNAME +$ " was born" $+$ " on " $+ BDATE$

The mechanism for resolving common expressions identifies $DNAME$ and " was born" as such and, instead of creating two different phrases, it creates one that combines both pieces of data:

$DNAME$ was born in $BLOCATION$ on $BDATE$

For the translation of the contents of a whole database containing multiple relations, we need to traverse the entire schema graph. This can be realized in several ways, e.g. with a simple DFS-like traversal starting from a central point of interest [9]. During this traversal, three possible structural patterns on the graph can be found: the *unary* pattern ($R_i - R_j$), the *join* pattern ($\binom{R_{i_1}}{R_{i_2}} > R_j$), and the *split* pattern ($R_i <\binom{R_{j_1}}{R_{j_2}}$). Translating multiple relations requires a careful and well-tuned combination of the above patterns for ensuring accurate and meaningful results.

For example, a split pattern $R_i <\binom{R_{j_1}}{R_{j_2}}$ may create clauses where the subject comes from the $R_i$ relation and the rest (e.g., the predicates) is based on the $R_{j_1}$ and $R_{j_2}$ relations. A possible translation may contain two different template clauses involving relations $R_i - R_{j_1}$ and $R_i - R_{j_2}$, respectively; an obvious challenge is to avoid repetition of the same information from $R_i$. (Observe that this case differs from the previous example, as in here, the repetition involves the whole information of $R_i$ and how that merges meaningfully with the translated content of the other two relations.) An appealing translation can be created by appropriately merging the above template clauses for producing a single template where the subordinate clauses are combined with a conjunctive term (e.g., "and"). For instance, the hypothetical schema $DIRECTOR \leftarrow MOVIES \rightarrow ACTOR$, where for some reason the translation has to be realized following the direction of the arrows, constitutes a split pattern. The straightforward translation would contain one phrase combining only the heading attributes of the involved relations, and then, phrases translating the content of the each individual relation. However, instead of getting a vapid narrative like:

"The movie M1 involves the director D1 and the actor A1. The director D1 was born in Italy. The actor A1 is Greek."

it would be more appealing to follow the aforementioned logic and get the following translation:

"The movie M1 involves the director D1 who was born in Italy <u>and</u> the actor A1 who is Greek."

As a more detailed example, assume that we want to translate contents of a subset of the graph depicted in Fig. 1 that contains relations $DIRECTOR$ and $MOVIE$ (this case resembles a sequence of unary patterns: $DIRECTOR - DIRECTED - MOVIE$). $DIRECTED$ participates in the translation process (as a node of the database schema graph) only for connecting the other two; none of its attributes contributes to the result, so it is not taken under consideration for the construction of the narrative. Therefore, conceptually, this case resembles a single unary pattern: $DIRECTOR - MOVIE$. Consider, for example, that the contents of this schema subset consist of three movies of director "Woody Allen". In this case, translation proceeds as follows. First, we construct the template clause that corresponds to relation $DIRECTOR$:

$DNAME$ was born in $BLOCATION$ on $BDATE$

The corresponding template clause for relation $MOVIE$ can be the following:

$TITLE +$ " (" $+ YEAR +$ ")"

Since relation $MOVIE$ may contain more than one tuple, we iterate over the above template for all the tuples. We proceed with the clause that corresponds to the relationship that connects $DIRECTOR$ and $MOVIE$. The template label of this relationship can be represented by the following:

"As a director, " $+ DNAME +$
   "'s work includes " $+ MOVIE\_LIST$

$MOVIE\_LIST$ contains two loops bounded by the arity of the movie tuples (similarly, arity of movie titles) and may be defined as:

$DEFINE\ MOVIE\_LIST$ as
   $[i < arityOf(TITLE)]$
      $\{TITLE[i] +$ " (" $+ YEAR[i] +$ ")," $\}$
   $[i = arityOf(TITLE)]$
      " and " $+ \{TITLE[i] +$ " (" $+ YEAR[i] +$ ")." $\}$

The result of applying all these steps, including instantiation of the template and the appropriate concatenations, on the database part that is relevant to "Woody Allen" in relation $DIRECTOR$ may be the following:

"Woody Allen was born in Brooklyn, New York, USA on December 1, 1935. As a director, Woody Allen's work includes Match Point (2005), Melinda and Melinda (2004), and Anything Else (2003)."

With a slightly different template, we could represent the same information using more than one clause:

"Woody Allen was born in Brooklyn, New York, USA on December 1, 1935. As a director, Woody Allen's work includes Match Point, Melinda and Melinda, Anything Else. Match Point was released in 2005. Melinda and Melinda was released in 2004. Anything Else was released in 2003."

The two pieces of text have some critical differences. The first one is more compact, does not have any overlaps, is *declarative*, and resembles genuine natural language. On the other hand, its creation is more complex and in more complicated cases may even be infeasible. For instance, if relation $MOVIE$ contains more attributes and multiple clauses are needed to describe them, then it is difficult to create such elegant result with the template method. The second piece of text is constructed in a *procedural* manner and consists of a coalescence of several simple sentences. This kind of synthesis is simpler to create and can be used to describe more complex database schema graphs. Automatically choosing between the two based on the characteristics of the database part concerned at any point is a great challenge.

Another observation concerns the text size. Although in principle, the approach outlined above works with databases of any size, translation of a database with a very large number of relations, attributes or tuples, will most likely lead to less meaningful or concise answers. Based on some limited interaction with potential users, it is clear that meaningful and interesting answers are short. Hence, an additional challenge is limiting the resulting text to the most interesting information. This can be realized either with structural constraints affecting the traversal of the database schema graph based on weights on its nodes and/or edges, or with some notion of ranking of the relations and tuples involved. The latter would force the most significant tuples to be presented first and the less significant tuples to be ignored according to appropriate constraints. Additionally, it is possible to have personalized settings (e.g., different heading attributes for relations or different weights on nodes and edges) in order to produce customized narratives for different users or user groups.

## 3. USER INTERACTION ELEMENTS

### 3.1 Motivation

Traditionally, the application of natural-language techniques to the front-end of an information systems environment has been one-directional: from NL descriptions to queries production. In this section, we examine the other direction as well: translation of queries into narratives.

Moving temporarilly away from the schema of Fig. 1, consider the following schema with two tables: $EMP(eid, sal, age, did)$ and $DEPT(did, dname, mgr)$. Consider someone posing the following SQL query:

```
select e1.name
from   EMP e1, EMP e2, DPT d
where  e1.did=d.did and d.mgr=e2.eid
  and  e1.sal>e2.sal
```

There are several reasons why having the system provide a natural language interpretation of the query may be useful. Before the query is sent for execution, it may be nice for the user to see it expressed in the most familiar way, as verification that the query captures correctly the intended meaning. Seeing something like "Find the names of employees who make more than their managers" for the above query will be very helpful in making sure that this was indeed the user's original intention. The more complicated the query, the more important such feedback is.

In general, in any situation where explanation of queries is warranted, such textual interpretation may be very useful and effective. For example, when a query returns an empty answer, it is nice to know the parts of the query that are responsible for the failure. Similarly, when a query is expected to return a very large number of answers, it is useful to know the reasons, in case a rewrite would reduce the number significantly and would serve the user better.

Clearly, the same can be said about all other commands a user may give to a database system. Insertions, deletions, and updates, especially those with complicated qualifications or nested constructs, will benefit from a translation into natural language. Likewise for view definitions and integrity constraints, which borrow most of their syntax from queries. Also, although here we focus on SQL, similar arguments can be made about Relational Algebra queries, RDF queries in SPARQL or RQL, even Datalog programs, and others. One can claim that novice users may benefit by textual specification of even queries posed by filling out a form. Especially for large forms, where a user is likely to not know the underlying semantic connections among the fields presented in the form, a textual explanation may come in handy.

Needless to say, offering the functionality described above is not trivial for complicated queries and other commands. Part of the complexity lies with the fact that there are several alternative expressions of a query in a formal language that are equivalent, based on associativity, commutativity, and other algebraic properties of the query constructs. Capturing the query elements in the right order so that the corresponding textual expression is natural and meaningful independent of the way the user has expressed the query is not straightforward. Similarly, expressing queries with complex embeddings or aggregations is hard.

### 3.2 Graph-based Query Representation

To abstract away the details of the above difficulties and translate queries in a generic fashion, a graph-based model can be used. Such a model is useful since (a) it is generic enough to capture queries expressed in different languages, (b) it can be visualized easily, so it offers an additional opportunity for studying queries, and (c) it can be annotated appropriately with suitable templates, so a narrative can then be created using appropriate parsing techniques.

Unfortunately, the graph model presented in 2.2 cannot capture the full expressive power of SQL or other common query languages, since queries cannot be always expressed in terms of subgraphs of the database schema graph. (This will become clearer with the examples of Section 3.3.) Hence, an extension to the schema graph model is needed. Next, without loss of generality, we sketch the necessary characteristics of such a model, relying on well-defined standard techniques, i.e., UML notation.

The schema graph representing the query, i.e., the *query graph*, comprises a set of nodes representing relations involved in the query, along with additional nodes representing other query functionality, e.g., group-by or order-by semantics. Each relation $R$ participating in a query $Q$ can be considered as a parameterized class (see Fig. 2), where the parameter is an alias for the relation, *relation_alias*, corresponding to the tuple variables of the relation. Such alias is useful when multiple instances (tuple variables) of a relation participate in query $Q$. We extend the traditional definition of a class, and we consider that it comprises four parts. The first part specifies the name of the class, i.e., of the relation, *relation_name*, and it is tagged with the label $<<FROM>>$. The second part involves those attributes of relation $R$ that participate in query $Q$ and is tagged with the label $<<SELECT>>$. The elements of this part follow the form: $relation\_alias.relation\_name.attribute$: $attribute\_alias$. The third part involves query constraints and is tagged with the label $<<WHERE>>$. The last part comprises grouping constraints and is tagged with the label $<<HAVING>>$. The grouping attributes and the ordering of the relation are captured by two UML notes, $<<GROUP\ BY>>$ and $<<ORDER\ BY>>$, respectively.

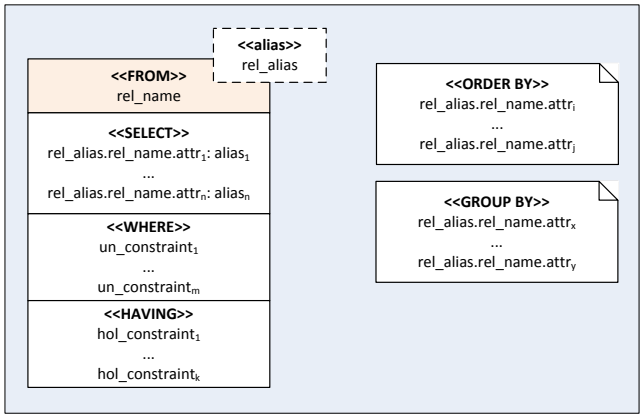

Figure 2: Schematic representation of a relation participating in a generic SQL query

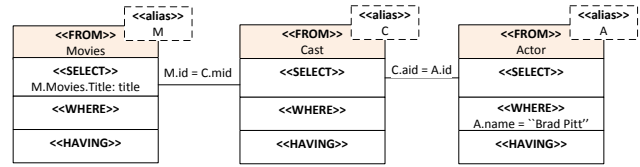

Figure 3: A simple path query ($Q1$)

The edges of the query graph can be generic join edges expressing arbitrary join conditions, projection edges connecting the relation with its attributes contained in the query, edges that connect the relation with elements of its 'WHERE' and 'HAVING' parts and with the 'GROUP BY' and 'ORDER BY' notes, and edges that connect the inner and outer parts of a nested query. Similarly to the annotation of the database graph (see Section 2.2), template labels can be assigned to those edges.

Hence, again, the translation can be realized in terms of traversing the query graph; however, this is nontrivial for all query types. To illustrate the difficulty that such translations represent, we present below several SQL query examples of escalating complexity. All queries correspond to the schema depicted in (Fig. 1) and, whenever possible, we show the respective query graphs considering only relation nodes and join edges. We do not aim at demonstrating a complete solution; rather, our goal is to pinpoint the existing challenges, to present a categorization of queries w.r.t. the effort needed for their translation into narratives, and to show that a graph-based approach similar to the one used for the translation of the content is feasible for a large variety of queries.

## 3.3 Interesting Cases

### 3.3.1 Path Queries

These are simple queries, whose graph representation is a path on the schema graph. Algebraically, these are select-project-join (SPJ) queries with at most two joins per relation and only one instance (tuple variable) per relation.

**Query 1.** Consider a simple query, $Q1$, which involves the relations $MOVIES$, $CAST$, and $ACTOR$ as follows:

**select** m.title
**from** MOVIES m, CAST c, ACTOR a
**where** m.id = c.mid **and** c.aid = a.id
  **and** a.name = 'Brad

The graph-based representation of $Q1$, a simple path, is depicted in Fig. 3. The joins are based on foreign key (FK) relationships and are depicted as edges connecting the relations involved. Each class representing a relation contains the appropriate attributes or constraints. Like all path queries, $Q1$ can be translated relatively easily: using translation mechanisms similar to those for database contents (see Section 2.2) and following a simple DFS-like traversal of the query graph, we can compose sentences such as:

  "Find the titles of movies where the actor Brad Pitt plays".

Furthermore, using more elaborated translation techniques, when more complex template labels are available, we can produce even more natural phrases, like:

  "Find movies where Brad Pitt plays"

Such phrases are created when the heading attribute is replaced by the conceptual meaning of the relation –e.g, 'title'→'movies' and 'name'→'actor'– or by the tuple variable –e.g., 'name'→'actor' →'Brad Pitt'.

### 3.3.2 Subgraph Queries

These are somewhat more difficult queries, whose graph representation is any (acyclic) subgraph of the schema graph (not necessarily a path). Algebraically, these are still select-project-join (SPJ) queries with only one instance (tuple variable) per relation, but no constraint on the number of joins a relation participates in.

**Query 2.** Consider the following query, $Q2$, which involves a large number of relations interconnected via FK join relationships:

**select** a.name, m.title
**from** MOVIES m, CAST c, ACTOR a,
    DIRECTED r, DIRECTOR d, GENRE g
**where** m.id = c.mid **and** c.aid = a.id
  **and** m.id = r.mid **and** r.did = d.id
  **and** m.id = g.mid **and** d.name = 'G. Loucas'
  **and** g.genre = 'action'

The graph-based representation of $Q2$ is depicted in Fig. 4. Given that subgraph queries do not deviate at all from the underlying database schema, they can again be translated using translation mechanisms similar to those for database contents (see Section 2.2). For example, with appropriate templates, $Q2$ can be translated into:

  "Find the actors and titles of action movies directed by G. Loucas"

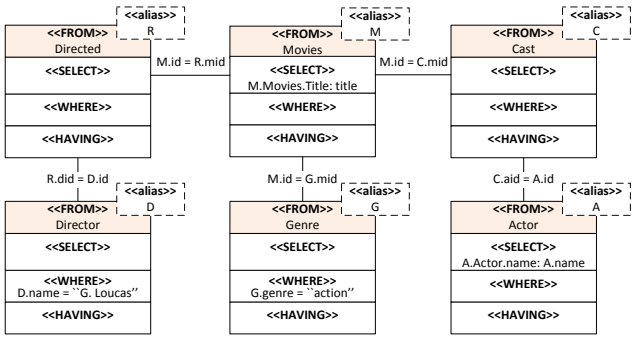

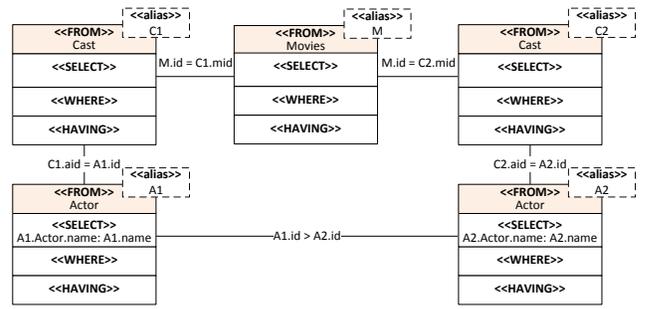

Figure 5: A multi-instance query ($Q3$)

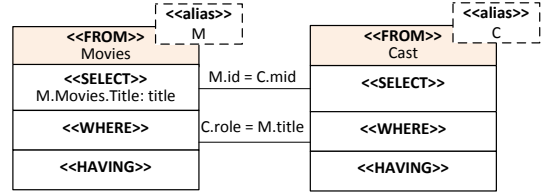

Figure 4: A more complex subgraph query ($Q2$)

### 3.3.3 Graph Queries

At the next level of difficulty, one finds queries that can still be represented by a graph that is a (possibly cyclic) subgraph of the schema graph or at least of an extension of it; in the sense that it may contain multiple instances of an existing node. Algebraically, these are all select-project-join (SPJ) queries with no restrictions.

**Query 3.** Consider the following query, $Q3$, which involves two instances (tuple variables) of some relations:

    select a1.name, a2.name
    from MOVIES m, CAST c1, ACTOR a1,
         CAST c2, ACTOR a2
    where m.id = c1.mid and c1.aid = a1.id
      and  m.id = c2.mid and c2.aid = a2.id
      and  a1.id > a2.id

The graph-based representation of $Q3$ is depicted in Fig. 5. It is still based on the schema graph, but now it has pieces of it repeated in multiple copies. Assuming that the translation templates are repeated on all of these copies and exist for the relevant non-FK joins, the database content translation techniques we have summarized above (in Section 2.2) could certainly be applied to generate a relevant piece of text. However, the result would be quite unnatural, even in the best case:

*"Find the name of an actor who has played in a movie, and the name of another actor who has played in the movie, and the id of the first actor is larger than the id of the second actor"*

Generating a natural sentence for this query requires that whole parts of the query graph be translated into individual phrases, essentially moving away from local template labels, that are associated with single attributes and assigning them to larger schema/query parts. How to define such template labels, whether or not there are any general patterns that should be followed, and so on, are open issues that require investigation. Based on how a human would translate query $Q3$, which is given below, it seems that identifying an effective approach is nontrivial:

*"Find pairs of actor who have played in the same movie"*

**Query 4.** Similar difficulties arise when the query graph contains cycles. Again, non-local template labels are needed, including some for non-FK joins, to capture a query naturally. Query $Q4$ below belongs to this category:

Figure 6: A cyclic query ($Q4$)

    select m.title from MOVIES m, CAST c
    where m.id = c.mid and c.role = m.title

Obtaining its translation from its query graph (Fig. 6) is a challenge:

*"Find movies whose title is one of their roles"*

### 3.3.4 Non-Graph Queries

Non-graph queries are those that cannot be represented on top of the schema graph or an expanded version of it with multiple copies of some of its parts. Algebraically, these are queries that involve operators other than select-project-join, or at least expressing them with just these operators is not obvious from their syntax. There are essentially two types of such queries: nested and aggregate queries.

Nested queries can also be classified into two categories: those that do have a flat (SPJ) equivalent and those that do not, as exemplified below. Each category presents its own challenges for translation.

**Query 5.** Consider the following nested query, $Q5$, where **in** is the only nesting connector:

    select m.title from MOVIES m
    where id in (
      select c.mid from CAST c
      where c.aid in (
        select a.id from ACTOR a
        where a.name = 'Brad Pitt'))

Clearly, query $Q5$ has a flat equivalent described in query Q1:

    select m.title from MOVIES m, CAST c, ACTOR a
    where m.id = c.mid and c.aid = a.id
      and  a.name = 'Brad Pitt'

Hence, the translation desired would be similar to the following:

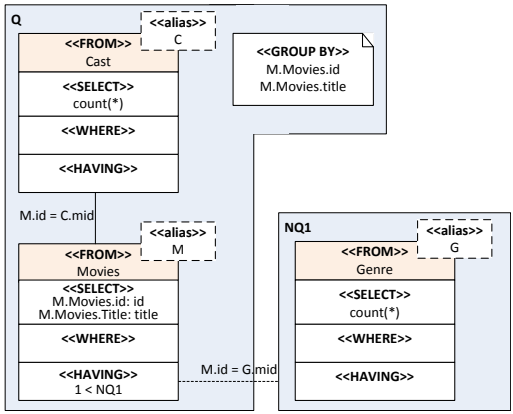

Figure 7: An aggregate query ($Q7$)

*"Find movies where Brat Pitt plays"*

Obtaining this from the original form is almost impossible, while it is straightforward to obtain from the flat form of the query. Hence, identifying equivalent query forms is important and receives new life as a problem when motivated by translatability principles,

**Query 6.** Consider the following nested query, $Q6$, where connectors other than **in** take part in the nesting:

**select** a.title **from** MOVIES a
**where not exists** (
　**select** ∗ **from** GENRE G1
　**where not exists** (
　　**select** ∗ **from** GENRE a2
　　**where** a2.mid = m.id))

Translating the above rather intertwined query is clearly a complex task, as ideally, one would want something rather short for its translation:

*"Find movies that have all genres"*

**Query 7.** Aggregate queries present similar challenges, since one cannot rely on the underlying graph. The magnitude of the problem is just shown with the query $Q7$:

**select** m.id, m.title, **count**(∗) **from** MOVIES m, CAST c
**where** m.id = c.mid
**group by** m.id, m.title
**having** 1 < (**select count**(∗)
　　　　　**from** GENRE g
　　　　　**where** g.mid=m.id)

A graphical representation of $Q7$ is depicted in Fig. 7. Observe that the nested part of the query is represented graphically as an additional query ($NQ1$). (Again, for simplicity most of the edges are not depicted.) That facilitates a procedural translation of the query; however, the challenge is to create a declarative one:

*"Find the number of actors in movies of more than one genre"*

### 3.3.5 Impossible Queries

Even in the most difficult of the cases mentioned so far, one could imagine ways to be explored in order to identify the appropriate translation techniques. There are some, however, that appear simply "impossible". These are queries whose semantics are not derivable from the query-graph representation and require higher-order languages, which are a very challenging to use every day.

**Query 8.** In query $Q8$, the semantics is unclear. Syntactically, one sees a standard aggregate query, but in reality, it is the **count** aggregate that implies **all** and dominates the query.

**select** a.id, a.name
**from** MOVIES m, CAST c, ACTOR a
**where** m.id = c.mid **and** c.aid = a.id
**group by** a.id, a.name
**having count**(**distinct** m.year) = 1

Hence, it is not obvious how to produce the correct narrative:

*"Find actors whose movies are all in the same year"*

**Query 9.** In query $Q9$, the semantics is unclear as well, but in a rather different way. This time, syntactically, one does see the **all** connector as the main challenge, but the expression '= all' will have to be interpreted as 'earliest' in this case, which is very difficult to obtain.

**select** a.name
**from** MOVIES m, CAST c, ACTOR a
**where** m.id = c.mid **and** c.aid = a.id
　**and** year <= **all** (
　　**select** m1.year
　　**from** MOVIES m1, MOVIES m2
　　**where** m1.title = m.title **and** m2.title = m.title
　　　**and** m1.id != m2.id
　)

Consequently, it is very difficult to produce the following text:

*"Find the actors who have played in the earliest versions of movies that have been repeated"*

For a system to recognize that a good way to express the meaning of these relatively simple queries is with phrases like the above is nontrivial. Identifying the correct use of pronouns is one source of difficulties. Another one is related to whether or not the description will be declarative (as in the above two examples) or procedural, i.e., whether it will just specify what the query answer should satisfy or also the actions that need to be performed for the answer to be generated. The former is always desirable, but for complicated queries, the latter may be the only reasonable approach. Identifying the complexity point where this becomes the case, however, is far from understood, and work must be done on this.

## 4. RELATED WORK

Earlier interaction of databases and natural language processing has focused mainly on the opposite direction of the one considered in this paper. Several works are presented concerning NL Querying [5], NL and Schema Design [12], NL and Database interfaces [1, 6], and Question Answering [10]. Hence, as far as we are aware of, related literature on NL and databases has focused on totally different issues such as the interpretation of users' phrasal questions to a database language, e.g., SQL, or to the automatic database design, e.g., with the usage of ontologies [11]. Several recent efforts use phrasal patterns or question templates to facilitate the answering procedure [6, 10].

In earlier work, we have studied the problem of translating small databases or query answers under certain constraints with promising results [8, 9]. In that work, the translation involves databases with content, i.e., the translation is performed at the data level. In this work, we investigate issues related to the translation of queries too, thus, we mainly work at the schema level. In a previous work, we have discussed the usefulness of translating SQL queries into narratives [4]. Here, we elaborate on that, we examine the space of the problem, and we discuss useful directions for reaching such goal.

## 5. CONCLUSIONS

In this paper, we have looked into the intersection of the Database and Natural Language Processing areas and have outlined several interesting problems that arise when one attempts to translate database elements into natural language elements. We have discussed the translation of both database contents and queries. Translating database content is not an easy task, mainly because it is not straightforward to choose the appropriate schema constructs and data items for composing a concise and meaningful narrative. Apart from such structural problems, identifying the right linguistic constructs, introducing pronouns where appropriate, and synthesizing everything to produce a natural end result are equally complex. On the other hand, translating database content to natural language is simpler from translating queries, as the extent of alternative equivalent expressions of schemas and data is much narrower than that of queries.

Additionally, we have offered example applications that indicate the practical usefulness of the problem, have identified several categories of database elements whose translation into text would be useful, and have briefly described some of the technical challenges that need to be addressed in the future. We hope that researchers will take up this type of problems and help to push this interesting area forward.

## 6. ACKNOWLEDGEMENTS

We are grateful to Georgia Koutrika for sharing her ideas with us and for commenting on several versions of this paper.